%% file: main.tex
\documentclass[12pt]{article}
\usepackage{sbc-template}
\usepackage{graphicx,url}
\usepackage[export]{adjustbox}
\usepackage{textcomp}
\usepackage{amsmath}
\usepackage[utf8]{inputenc}  
\usepackage{subfigure}
\usepackage{makecell}
\usepackage{changepage}

\usepackage{graphicx}
\usepackage{graphics}
\usepackage{footnote}
\usepackage{tablefootnote}
\usepackage{environ}
\usepackage{xurl}
\usepackage{algpseudocode}
\usepackage{amssymb}
\usepackage{xcolor}
\usepackage[ruled, linesnumbered]{algorithm2e}
\usepackage{microtype}
\usepackage{comment}

\newcommand{\mpolka}{\textit{M-PolKA}\xspace}
\newcommand{\mpolkaint}{\textit{MPolKA-INT}\xspace}

\NewEnviron{NORMAL}{%
    \scalebox{2}{$\BODY$} 
} 
 
\NewEnviron{SHORT}{%
    \scalebox{1}{$\BODY$} 
}

\newcommand{\nome}[0]{\textit{MM-INT}\xspace}

\title{{\nome: Telemetry in Programmable Switches with Multiple Queues using Source-based Multipath Routing}}

\author{Mateus N. Bragatto\inst{1}, João Paulo M. Clevelares\inst{2}, Cristina K. Dominicini\inst{3}, \\Rodolfo S. Villaça\inst{2}, Fábio L. Verdi\inst{1}}

\address{Department of Computing (DComp) -- Federal University of São Carlos (UFSCar)\\
  Sorocaba/SP, Brazil.
    \vspace{-0.3cm}
    \email{mateusbragatto@estudante.ufscar.br, verdi@ufscar.br}
\nextinstitute
  Department of Informatics (DI) -- Federal University of Espírito Santo (UFES)\\
  Vitória/ES, Brazil.
    \vspace{-0.3cm}
    \email{joao.clevelares@edu.ufes.br, rodolfo.villaca@ufes.br}
\nextinstitute
  Federal Institute of Espírito Santo (IFES) -- Serra Campus\\
  Serra/ES, Brazil.
    \vspace{-0.3cm}
    \email{cristina.dominicini@ifes.edu.br}
}

\begin{document} 

\maketitle

\input{sections/resumo}

\input{sections/intro}

\section{Theoretical Background and Research Problem Positioning} \label{sec:teorica}

This section is divided into two parts: i) a description of how metric collection works in programmable networks using INT, and how multiple queues affect this collection; and ii) how MPolka-INT works.

\subsection{Telemetry using INT}
\label{INT}

Thanks to recent technological advances, the popularization of programmable network devices, and the P4 language, it is now possible for these devices to report the state of the network without control plane intervention \cite{McKeown:2019}. In this case, data packets can contain headers that are telemetry instructions, which can be used to collect performance data. The telemetry instructions are described in the INT data plane specification, which defines three modes of operation: INT-XD (eXport Data), INT-MX (eMbed Instructions), and INT-MD (eMbed Data).

In INT-XD mode, each device exports the metadata using INT instructions pre-configured in the flow tables, acting directly from the data plane to the monitoring system. In this mode, no modifications are made to the client traffic packets. In INT-MX mode, the source node creates INT instructions in the packet header, so that at each transit node, the INT instructions are read, the respective metadata is collected, and transmitted directly to the monitoring system. In this mode, small modifications are made to the data packets originated by clients, since telemetry instructions need to be inserted into the packet headers. In INT-MD mode, INT instructions and metadata are inserted into the packets at each hop in the network. This is the default operating mode defined by the INT specification and the one chosen for this work. To avoid confusion, INT-MD mode will be referred to as INT throughout this paper.

\begin{figure}[ht]
    \centering
    \includegraphics[scale=0.40]{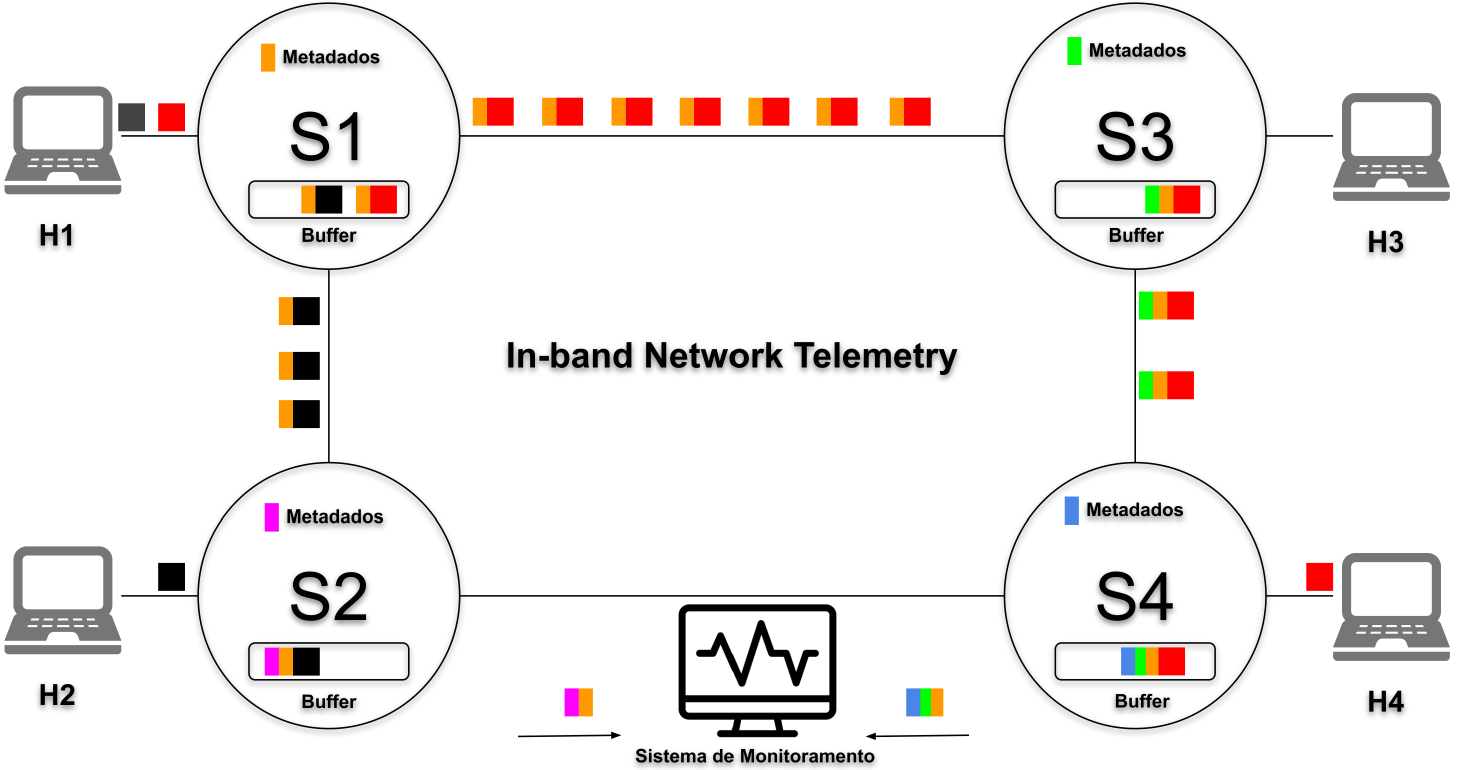}
    \caption{Example of a network with In-band Network Telemetry.}
    \label{fig:INT}
\end{figure}

Fig. \ref{fig:INT} illustrates the operation of INT in a network. The network consists of four end systems (\textit{H1, H2, H3}, and \textit{H4}) and four programmable \textit{switches} with support for INT telemetry (\textit{S1, S2, S3}, and \textit{S4}). Each \textit{switch} has a set of metadata of interest to the monitoring system, represented by rectangles in orange (\textit{S1}), magenta (\textit{S2}), green (\textit{S3}), and blue (\textit{S4}).
%
%
%
In the network, there are two data flows: one represented by red packets and the other by black packets. The red flow follows the path \textit{f1 = \{H1, S1, S3, S4, H4\}} in the network. The black flow follows the path \textit{f2 = \{H1, S1, S2, H2\}}. At each network hop, telemetry instructions in the data plane of the devices guide the collection and addition of metadata to the packets traversing each \textit{switch}. This process is repeated along the entire path, from the first \textit{switch} after the source to the last \textit{switch} before the destination. At the destination \textit{switch}, the metadata is extracted from the packet and forwarded to a network monitoring system. The original packet is then delivered to its destination.

One of the main attractions of using INT lies precisely in the level of granularity achieved, as the packets traversing the network carry information for the monitoring system. 
In addition to the modes described in the INT specification, other approaches for collecting metadata in programmable networks can be used. One of these is the possibility of using a flow of exclusive packets for network telemetry. These exclusive packets are called \textbf{probes} and are responsible for collecting telemetry metadata, thus not altering the data packets. The main advantage of this approach is to avoid fragmenting the data packets, as adding telemetry metadata to these packets could exceed the MTU (\textit{Maximum Transmission Unit}) of the link layer. The main disadvantage of this approach is the need to cover the entire network topology for sending the probes.
Typically, traditional solutions generate an excessive number of sondas to cover the entire network topology. In this sense, the use of \mpolkaint, as proposed in this paper, reduces the number of probes required for this task through the use of source-based routing and multicast trees, which will be detailed in Section \ref{MPolKA}.

\begin{figure}[ht]
    \centering
    \includegraphics[scale=0.35]{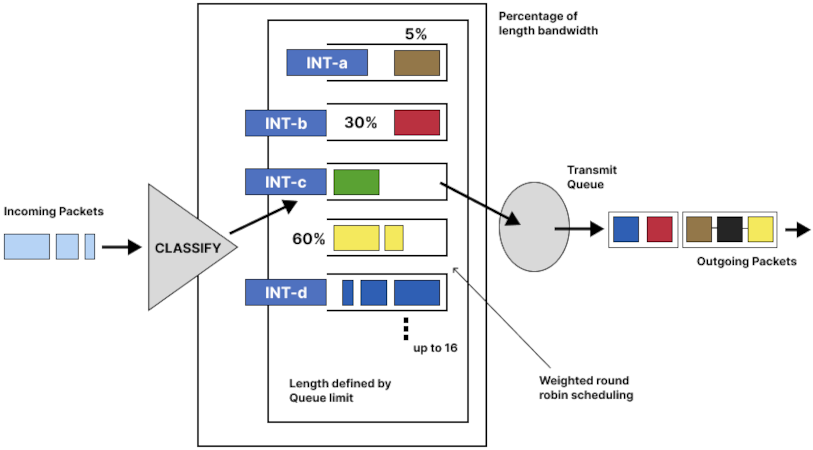}
    \caption{Multiple queues on an output port of a network device. Adapted from \cite{QueueDisciplines}.}
    \label{fig:queueDiscipline}
\end{figure}

Up to this point, it has been presented how INT telemetry operates to collect metadata from network devices. However, nothing has been said about the multiple queues present in these devices. Fig. \ref{fig:queueDiscipline}, extracted from \cite{QueueDisciplines}, shows in detail the internal operation of an output queue in a network device. In the figure, it can be observed that the packets entering the device are directed to different output queues before being forwarded to the link. Each output queue has its own occupancy rate, which dynamically changes as more or fewer packets arrive and depart from the queue. In the figure, the removal of packets from the queue occurs through a weight-based scheduling, where a queue with higher priority (more weight) will have its packets removed (sent to the output link) more quickly.

Furthermore, the same Fig. \ref{fig:queueDiscipline} essentially illustrates the default behavior of a network device. However, considering a traditional programmable device, capable of being monitored via INT telemetry, some observations must be made. First, an INT probe is only capable of collecting metrics (metadata) from the queue in which it is enqueued. In other words, for each queue on the port associated with the outgoing link, a different INT probe must be generated so that the metrics of each queue along a path are collected. It can be observed in the figure the presence of multiple INT probes, represented in blue, one for each queue (INT-a, INT-b, etc.). Each probe illustrates the collection of metadata individually in its specific queue.

However, as already mentioned, this solution generates the need for a large number of probes to collect monitoring data from all the queues present on the outgoing port of a programmable device, making it unfeasible to collect telemetry from the entire network. 
The use of registers, provided by programmable devices, allows for the storage of telemetry data to be collected when a packet passes, and the telemetry data is maintained for subsequent packets. Thus, by defining specific registers, it is possible to store information from each of the queues, where reading and writing will occur only in the register that represents the specific physical port and logical queue that a given packet has passed through.


\subsection{\mpolkaint}
\label{MPolKA}

\mpolkaint \cite{mpolkaint} proposes a mechanism to perform network telemetry, exploring multiple paths, reducing overhead in both the control and data planes, eliminating redundancy in telemetry information, and replacing probe routing tables with source-based routing. To cover the nodes of interest in the network, the proposal uses the \textit{Multipath Polynomial Key-based Architecture} (\mpolka) routing solution \cite{mpolka}, as it is a source-based routing method with support for multipaths, where a route label representing a cycle or tree can be encoded in a topology-agnostic manner \cite{mpolka}.

The M-PolKA routing is based on the Residue Number System (RNS) and polynomial arithmetic using Galois fields of order 2, GF(2). In this scheme, at core nodes, the transmission states of the output ports are given by the remainder of the binary polynomial division (polynomial mod operation) of the packet's route identifier by the network device's node identifier. Its implementation in programmable switches leverages packet cloning mechanisms and the reuse of CRC (Cyclic Redundancy Check) hardware, which enables the polynomial mod operation.

\begin{figure}[ht]
\centering
\includegraphics[width=\textwidth]{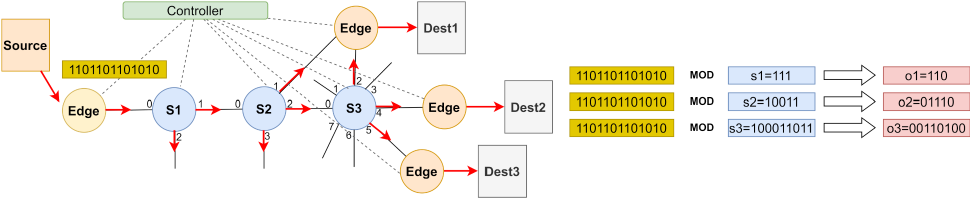}
\caption{Example of M-PolKA routing \cite{mpolka}}
\label{fig:mpolka}
\end{figure}

As shown in the example in Fig. \ref{fig:mpolka}, the \mpolka architecture consists of: (i) edge nodes (in yellow), (ii) core nodes ($s_1$, $s_2$, and $s_3$, in blue), and (iii) a logically centralized controller (in green). The routing relies on three polynomial identifiers: (i) \textit{nodeID}: a fixed identifier assigned to the core nodes by the controller during a network configuration phase; (ii) \textit{t\_state}: an identifier assigned to the transmission state of the output ports at each core node; and (iii) \textit{routeID}: a multipath route identifier, calculated by the controller using the RNS and incorporated into the packet by the edge nodes. In Fig. \ref{fig:mpolka}, for an example flow, after calculating the \textit{routeID} ($10101100101100$), the controller installs rules at the edge to incorporate this label into the packets of this flow. Then, each core node calculates its \textit{t\_state} by dividing this \textit{routeID} by its \textit{nodeID}. At $s_1$, the remainder of the \textit{routeID} ($10101100101100$) when divided by its \textit{nodeID} ($111$) is $110$. Thus, at $s_1$, ports $1$ and $2$ forward the packets to the next hop. Similarly, at $s_3$ (\textit{nodeID} $100011011$), the result of the \textit{mod} operation is $00110100$, and only ports $2$, $4$, and $5$ forward the packets to the next hop.

In the telemetry solution of MPolKA-INT, the controller first calculates the \textit{routeID}, using the \mpolka routing described earlier, to represent the \textit{multicast} tree for the monitoring probe’s route that will cover all the nodes of interest. Then, the generating node inserts the calculated \textit{routeID}, as well as the telemetry information from the \textit{switch}, into the packet header. After that, the probe traverses each \textit{switch}, collecting telemetry information. At each hop, the packet is cloned to all active output ports in the transmission state vector. To eliminate redundancy, the telemetry metadata related to that node is inserted only on the first output port.

\input{sections/related}

\section{INT Metadata Collection in Multiple Queues}
\label{multiQueue}

In Section \ref{INT}, specifically in Fig. \ref{fig:queueDiscipline}, a limitation of the INT-based telemetry monitoring, in its traditional form, was presented regarding situations involving multiple queues on network device ports. The limitation is that an INT probe can only collect telemetry metadata from the queue associated with the port through which the probe is passing. Therefore, the simplest and most naive solution would be to send a probe for each queue present on all the ports of every switch in the network. However, this solution is not viable due to the number of probes that would be required to cover all queues on all ports and all switches in the monitored network.

Thus, this paper presents \nome{}, a solution that adopts the use of registers to store telemetry metadata for all queues present in programmable devices through the P4 language, so that an INT probe collects such data directly from these registers. Since registers are part of the shared memory of the device, a single probe is capable of reading these registers and collecting telemetry from all queues and ports of this device.

\begin{figure}[h]
\centering
\includegraphics[scale=0.6]{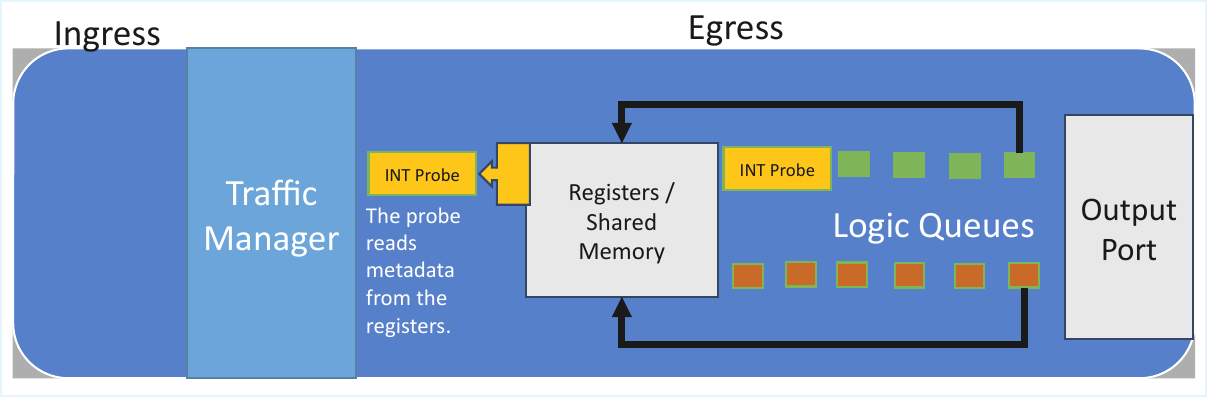}
\caption{Internal structure of each \textit{switch} for collecting telemetry metadata from multiple queues.
}
\label{fig:MultiQueue}
\end{figure}

Fig. \ref{fig:MultiQueue} illustrates the mechanism developed to support the \nome{} solution proposed in this paper. The figure presents a programmable \textit{switch} with two \textit{pipelines}: ingress and egress. Our solution is deployed only in the egress pipeline, after the \textit{Traffic Manager}, where the telemetry information for all the ports and queues of the device is located. The presented example has two queues on the physical port, one green and the other red. The mechanism works as follows: each data packet, before being removed from the queue to be transmitted to the outgoing link, invokes the calls to obtain the INT telemetry 

The telemetry data returned by these calls is stored in registers, with each queue having its own dedicated register. The INT probe, represented in yellow in the figure, is responsible for reading the stored telemetry data from the registers and appending it to its header. In \nome{}, the probe is capable of retrieving telemetry data from all ports and queues of a programmable \textit{switch}. After collecting the telemetry data, the probe is forwarded to the next \textit{switch}, following the path defined at the source. Furthermore, note that the INT probe will also be forwarded through a queue. However, in our solution, this probe does not invoke the INT calls to obtain telemetry data, since such data was obtained from the registers. In the figure, the INT probe was forwarded through the green queue. It is important to emphasize that, if the telemetry data were not stored in the registers, the INT probe would obtain the data only when being forwarded through a queue and would therefore be able to retrieve telemetry data from that queue only (in the figure, only from the green queue).


\label{sec:proposta}

\subsection{Detailed Analysis and Discussion of the Solutions}
\label{detalhandoSolucoes}



\begin{figure}[!h]
\centering
\includegraphics[width=0.8\columnwidth]{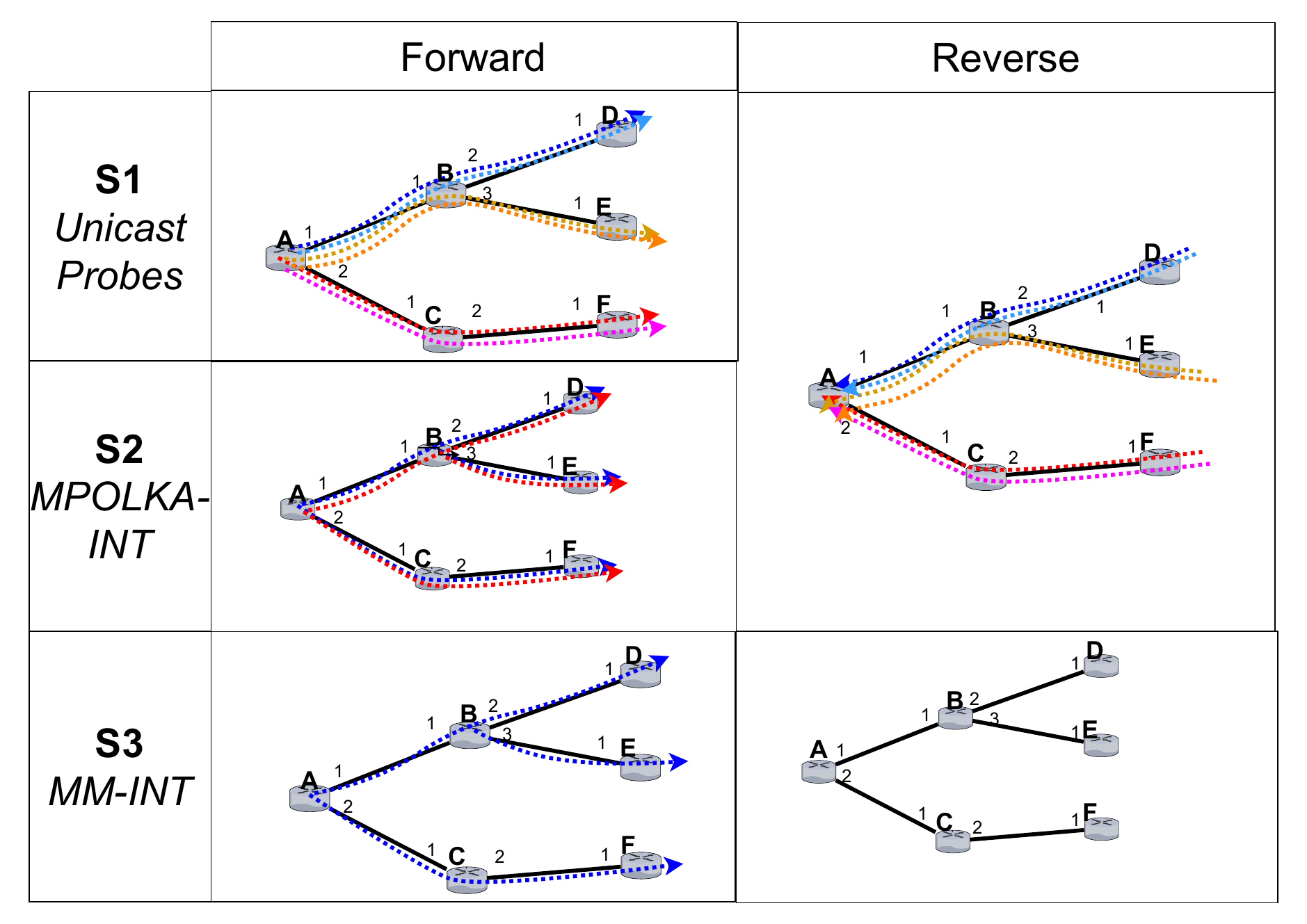}
\caption{Comparison with Solutions $S1$ and $S2$. Example with 2 queues per port.}
\label{fig:solucoes}
\end{figure}

To facilitate the understanding of the advantages associated with \nome{}, this section presents two traditional solutions found in the literature for telemetry data collection. These solutions will be compared with \nome{} in Section~\ref{sec:resultados}. In this context, the solutions found in the literature are those that do not use (or make little use of) mechanisms to reduce the number of INT probes sent through the network. Fig.~\ref{fig:solucoes} will be used to support this discussion.

\noindent{\textbf{Solution $S1$}}: this approach represents the simplest mechanism for telemetry data collection in the network, which sends probes to cover each existing queue on each port of the switches in the topology. The probes must be generated at the source (by some probe-generating host belonging to the monitoring system) and must be able to collect data from all queues on all ports of the equipment. It is important to recall that traditional INT probes collect information only from the queue associated with the output port through which they pass in the device. In Fig.~\ref{fig:solucoes}, the INT probes, represented by colored arrows, must be sent in both directions along the forwarding tree (forward and reverse), from the root node (switch A) to each leaf node (switches D, E, F), and also from each leaf node to the root, to cover the input ports on the switches along the forward path. In any tree, the number of INT probes can be obtained through the following equation: lf x nq x 2, where lf is the number of leaf nodes in the distribution tree and nq is the number of queues per physical port. In this example, we consider that the number of queues per port is the same for all switches in the network. In Solution S1, it is important to note the occurrence of duplicate collection, represented by the four colored arrows passing through the A-B (forward) and B-A (reverse) paths. These arrows represent the 4 repeated probes that pass through this link collecting the same metrics.

\noindent{\textbf{Solution $S2$}}: This solution introduces MPolKa-INT, meaning that only one probe is generated and sent in each direction of the multicast tree represented in Fig.~\ref{fig:solucoes}. Following the logic of MPolka-INT's operation, the probe must be cloned at each branch point to cover all the branches of the tree. However, since there are multiple queues per port, the probe must be cloned and recirculated for each queue to individually collect telemetry data from each queue associated with the output port of the probe. 
For example, consider the operation of switch B from Fig. \ref{fig:solucoes}. The INT probe arrives at switch B via port 1. This probe will be forwarded through a queue of port 2. However, before being forwarded, the probe will be cloned three times, one for each remaining queue (2 ports, 2 queues per port). Each cloned and recirculated probe will follow its respective queue, as shown in the figure, and then collect the telemetry metrics from the queue (through which it passed) associated with the output port of the probe. However, it can be observed that the simple application of MPolka-INT reduces the number of probes in the forward path, but in the reverse path, when the probe is sent from the leaf nodes, some duplicates still occur. First, the need to send one probe per queue remains. Additionally, in the example, the duplication of probes on the B-A link is evident by the number of arrows on this link along the reverse path. Therefore, Solution $S2$ reduces, but does not eliminate, the duplication of INT probes when compared to Solution $S1$.

\noindent{\textbf{Solution $S3$}, \nome{}: Fig.~\ref{fig:solucoes} also illustrates Solution $S3$, which makes use of \nome{}, where only one probe is generated at the source and will be cloned only when necessary, at the branches of the topology (branches of the multipath forwarding tree). As can be observed in Fig.~\ref{fig:solucoes}, the probes are generated only in the forward direction, since it is possible to collect telemetry data from all queues and ports of the same switch by reading the internal registers implemented in the solution. Thus, the need to send probes in the reverse direction is eliminated, and there is no duplication of probes on any link. In summary, with \nome{}, a monitoring system is capable of collecting telemetry data from all queues associated with all ports of a switch using a multipath tree with source routing, drastically reducing the number of probes needed to cover an entire topology, completely eliminating probe duplication in the network.





\subsection{\nome{} Implementation}
  
%
In the following examples, consider that by default a packet will always be sent to queue 0 of the output port of the programmable switch. In the parsing stage, the packet’s header field \textit{etherType} is used to check whether the packet is a telemetry probe (TYPE\_SR) or a data packet (TYPE\_IPV4). A data packet is forwarded using traditional routing tables based on the destination IP address. If the TOS field is not set to the value 55 (a configuration chosen for a telemetry packet), it is considered a data packet and queue selection is then performed. Based on the selection, telemetry data related to that queue is written to the corresponding registers.

The INT probe is forwarded in the network using the M-PolKA forwarding mechanism, described in Section \ref{MPolKA}. 
Thus, in the parsing stage for an INT probe, the value of the \textit{routeID} is obtained as the identifier of the multipath route to be traversed, in addition to extracting the INT header stack. 
After executing the pipeline, the packet is cloned to all active output ports in the transmission state vector (\textit{bit} 1). To eliminate redundancy in telemetry information, we use the \textit{resubmit} feature to modify the header of a cloned packet and insert telemetry metadata only into the first active output port.

\section{Experimentation and Evaluation of \nome{}}\label{sec:resultados}

The evaluations were carried out in a virtualized environment using Mininet. One of the first contributions in this regard was to adapt the Mininet emulator to support multiple queues, modifying the necessary files both in the architecture (\textit{v1model}) and in the compiler (\textit{p4c}). Although the BMv2 implementation supports up to eight logical queues per physical port, there was no version of Mininet that integrated BMv2 with this support.

This version is publicly available on GitHub\footnote{https://github.com/dcomp-leris/p4-multiqueue.} with a pre-configured virtual machine.

The topology used for the functional and comparative evaluation of \nome{} is shown in Fig.~\ref{fig:solucao1}. It is important to highlight that, for demonstration purposes, this topology is already in a tree format\footnote{\nome ~works on any topology in tree format or in a generic topology that is converted into a tree for routing purposes (e.g., a spanning tree). Due to space constraints, we will analyze only one topology in this paper.}. For this article, we configured two logical queues per port (\textit{queue 0 and queue 1}). The implementation was done using the P4 language and all artifacts for experiment reproduction are also available on GitHub\footnote{https://github.com/dcomp-leris/MM-INT.}.


\subsection{Functional Evaluation of the \nome{} Implementation}

The first experiment shows that the implementation of \nome{} is functional. The simplest way to demonstrate this was to generate heavy traffic in all directions between hosts connected to the leaf nodes of the topology (not shown in Fig. \ref{fig:solucao1}), so that all queues would become occupied. In parallel, INT probes were generated to follow the tree according to the multipath routes defined at the source by MPolka-INT. A Python application was developed to act as an INT probe collector on the edge switches (\texttt{SW1}, \texttt{SW3}, \texttt{SW4}, and \texttt{SW7}). Fig. \ref{fig:funcionamento} shows the queue occupancy, in number of packets, for switch \texttt{SW1} only. For the other switches, the graphs are similar and are not shown due to space limitations. Note that the figure shows the occupancy of the two logical queues on each physical port of switch \texttt{SW1}.

\begin{figure}[h]
\centering
\includegraphics[width=1\columnwidth]{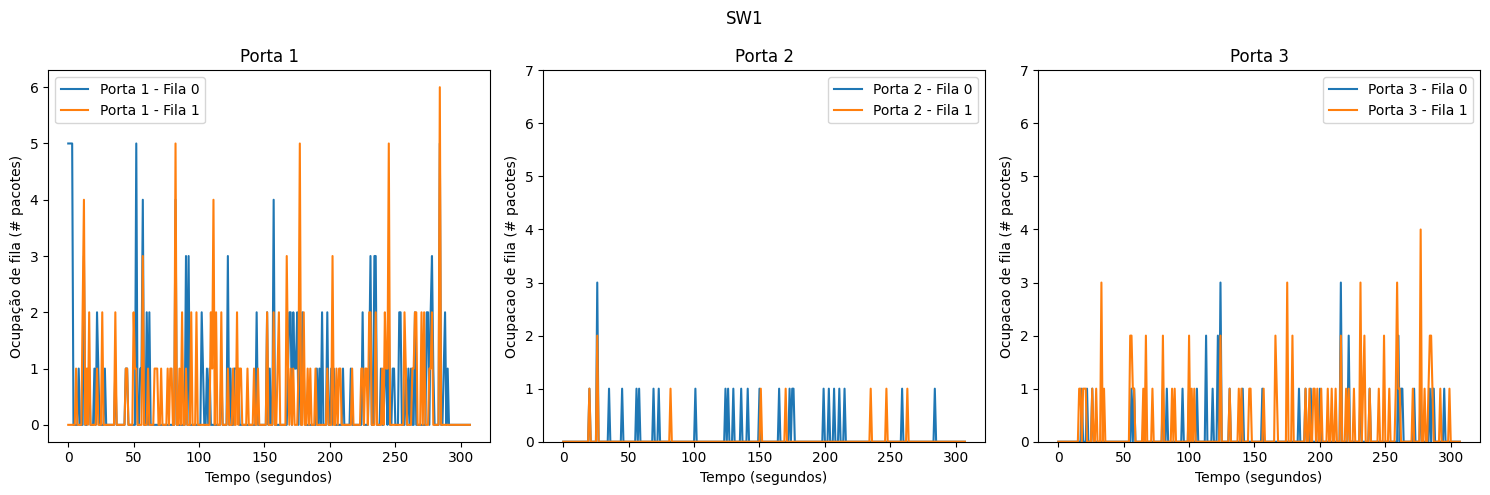}
\caption{INT probe collection performed on switch \texttt{SW1}.}
\label{fig:funcionamento}
\end{figure}

\subsection{Performance Evaluation of \nome}

Next, after ensuring that the implemented solution is working, we analyze the number of probes sent for solutions $S1$, $S2$, and $S3$ (\nome). Due to space constraints, the topology presented in Fig. \ref{fig:solucao1} illustrates the sending of probes only for Solution $S1$, in the forward direction, from the root node (\texttt{SW1}) to the leaf nodes. Therefore, one can observe the arrows (probes) leaving the switch \texttt{SW1} towards the switches \texttt{SW1}, \texttt{SW3}, \texttt{SW4}.

In this topology, considering Solution $S1$, 6 probes are required for the forward direction and 6 for the reverse. In the forward direction, we have 2 probes (1 for each logical queue) leaving \texttt{SW1} towards \texttt{SW3}, 2 probes leaving towards \texttt{SW4}, and 2 probes leaving towards \texttt{SW7}. In the reverse direction, from the leaf nodes to the root, the process is repeated, resulting in 6 more probes. Duplicates occur in both forward and reverse directions, with 2 probes on the link \texttt{SW1}-\texttt{SW2} and another 2 on the link \texttt{SW2}-\texttt{SW6}, totaling 4 duplicated probes in the forward direction and another 8 duplicated probes in the reverse, amounting to 12 repeated probes. It is important to highlight that the amount of duplicated probes depends on the topology. Other tree topologies will be analyzed in future work.

Considering Solution $S2$, the number of probes sent is the same as in Solution $S1$, that is, 12 probes. One of the main advantages of $S2$, which already uses MPolka-INT, is that only one probe is generated by the probe generator node, and the cloning occurs only at the branching points, as explained in Section \ref{detalhandoSolucoes}. Thus, fewer probes traverse the links, since there is no duplication of probes in the forward direction, only in the reverse. The number of repeated probes in the reverse is the same as obtained in Solution $S1$ (8 probes).

\begin{figure}[h]
    \centering
    \includegraphics[scale=0.45]{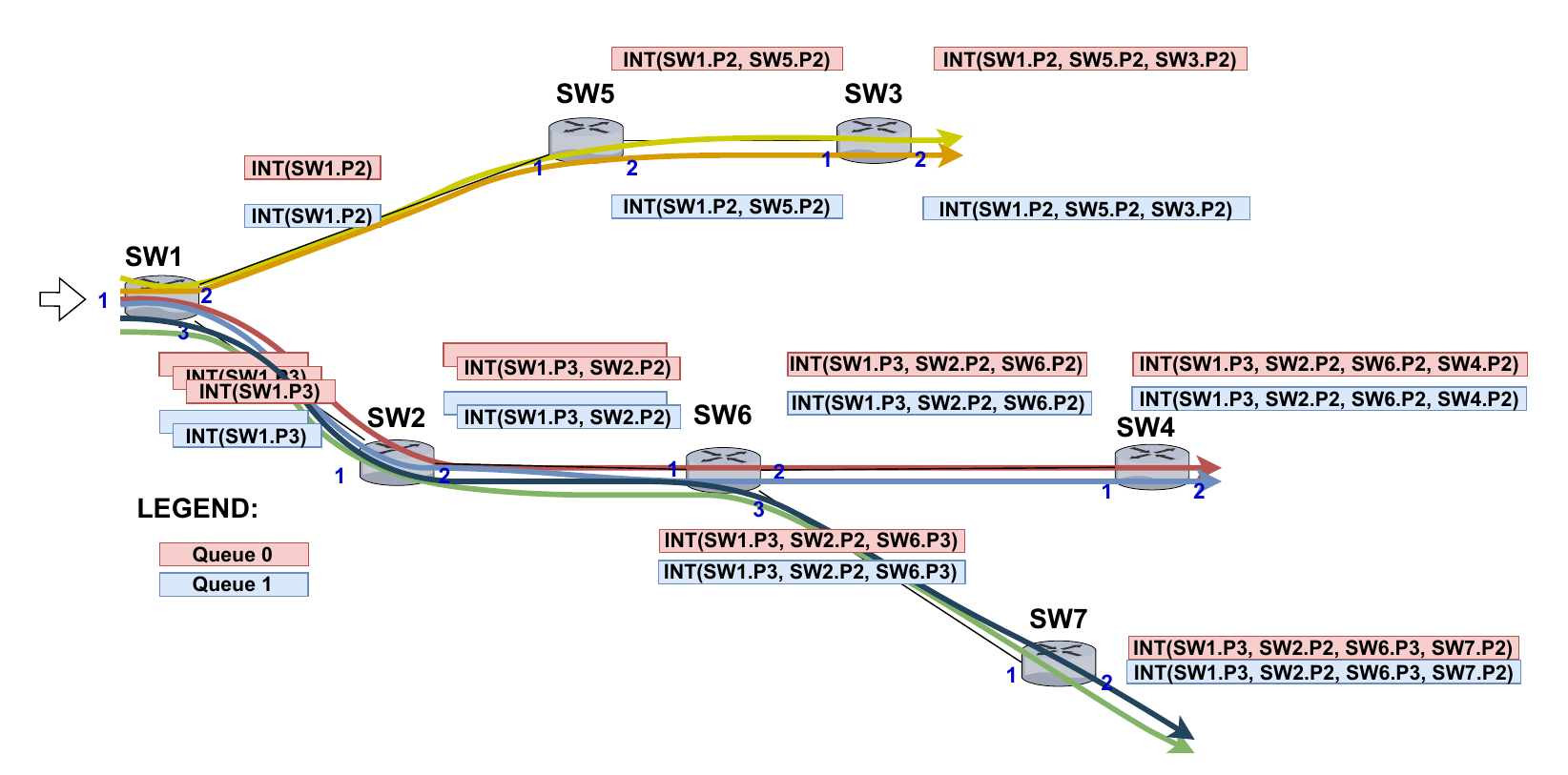}
    \caption{INT probe forwarding using Solution $S1$ (original INT). The figure represents only the probe departing from \texttt{SW1} to the leaf nodes (\texttt{SW3}, \texttt{SW4}, and \texttt{SW7}).}
    \label{fig:solucao1}
\end{figure}

In addition to the number of probes sent and the duplicates, other metrics must be analyzed to verify the advantages and disadvantages of \nome. For this purpose, the following metrics were also analyzed: probe size (in \textit{Bytes}), memory usage, and the amount of data (in \textit{Bytes}) required for the INT probes.

 

\begin{table}[ht] 
\caption{Evaluation results for the topology used in this article.} 
\scriptsize
\centering
\begin{tabular}{ |c|c|c|c|c|c| }
\hline
\textbf{Mechanism} & \textbf{Pkt. size (bytes)} & \textbf{\# probes} & \textbf{Mem. (in bytes)} & \textbf{\# dup.} & \textbf{Total bytes} \\ 
\hline
S1 & 61 & 12 & 0 & 12 & 2300\\  
\hline
S2 & 75 & 12 & 0 & 8 & 2174 \\  
\hline
\nome & 122 (2 ports) 154 (3 ports) & 3 & 64 (2 ports) 96 (3 ports) & 0 & 814\\
\hline
\end{tabular}
\label{tab:resultados}
\end{table}

Table \ref{tab:resultados} presents the obtained results. As can be observed, \nome{} uses a small amount of switch memory, since it needs to store telemetry information from the logical queues in the registers (stateful). Solutions $S1$ and $S2$ do not need to store any state in registers, and therefore do not consume memory space. However, the amount of memory used by \nome{} is negligible, and even in a switch with 32 physical ports and 128 logical ports per physical port (Tofino2), around 65KB would be required. Considering that this switch has 40MB of shared memory, we understand that the solution is scalable in terms of memory usage. On the other hand, the size of the INT probe for \nome{} in this case can reach approximately 65KB, as it embeds the information collected in these registers, which exceeds the Maximum Transmission Unit (MTU) limit of 1500 Bytes in an Ethernet network, resulting in fragmentation.

\nome{} stands out when analyzing the number of probes sent and duplicated, as it sends 4x fewer probes compared to the other evaluated solutions, without generating duplicate probes. Due to the smaller number of probes, each packet in MM-INT is able to collect more telemetry information in a single transmission. Therefore, when analyzing the packet size, \nome{} requires more Bytes to carry the telemetry metadata, and this size depends on the number of ports and queues per port. Depending on the packet size, the MTU of the link layer may be exceeded, causing fragmentation. In such cases, one strategy to limit the size of the collection packet is to selectively define what each probe is allowed to collect. It is possible to define which ports and metadata will be collected by each probe, as presented in existing solutions in the literature \cite{DINT-PINT,PINT,SEL-INT}.

Even though \nome{} has a larger telemetry packet, when analyzing the total amount of Bytes used for data collection in this topology, it is observed that \nome{} requires a much lower amount of Bytes compared to the other two solutions analyzed in this paper, approximately 2.8x fewer Bytes. Thus, even though \nome{} requires a larger header for telemetry, in the overall aggregate, less network bandwidth is used since it sends significantly fewer probes (4x reduction) and does not generate duplicates.





\subsection{Prototype Limitations}
\label{sec:limit}

The solution presented in this work proved to be feasible and functional when compared to traditional solutions, especially considering that it is a prototype in its early stage. However, there are some limitations that need to be highlighted, which will be the subject of future research and development efforts related to \nome:

\begin{itemize}

    \item The solution assumes that the number of logical queues per port is the same on all devices.

\item The INT packet size is larger, and may not scale if there are many physical ports and logical queues on the devices.

\item There is some delay between the telemetry information stored in the registers and the moment the probe passes through the \textit{switch}.

    \item The implementation
may undergo changes when applied to a physical device such as the Tofino \textit{switch}. Specifically, the issue of reads and writes performed on the registers and the number of pipeline execution stages will require adaptations to be supported in the \textit{TNA (Table Type Architecture)}.

\item Fragmentation caused by using \nome may occur due to the limitation imposed by the MTU. This point suggests a more thorough analysis regarding the limit related to the number of ports and queues per \textit{switch} collected, as well as the number of \textit{switches} a single probe could collect from.



\end{itemize}

\input{sections/conclusao}

\input{sections/ack}

\bibliographystyle{sbc}
\bibliography{sbc-template}

\end{document}

%% file: sections/resumo.tex
\begin{abstract}

This article emphasizes the importance of queues associated with the ports of switches in network monitoring. Traditionally, data collection about these queues is done using programmable data planes and telemetry based on INT (In-band Network Telemetry) probes, assuming there is only a single queue per output port. The MM-INT (Multiqueue Multicast - INT)
 is a solution that utilizes registers to store data from all queues and ports, enabling the efficient collection of monitoring information. The MM-INT avoids probe overload and employs the origin-based routing mechanism and multicast trees for the probes. The results demonstrate significant reductions in the number of probes sent compared to other traditional solutions found in the literature.
 
\end{abstract}

%% file: sections/intro.tex
\section{Introdução} \label{sec:introdução}

In network switches, queues are associated with the device's ports and represent an important source of monitoring information, as they can indicate the level of network congestion \cite{Kim:2018}. Moreover, their occupancy level can affect other metrics, such as response time, transmission delay, and jitter. However, collecting data about the queues in real time has always been a high-cost task due to the overhead generated in the network and the devices \cite{McKeown:2019}. Nevertheless, this scenario has changed in recent years thanks to the popularization of programmable data planes, with the P4 language, and monitoring with in-band telemetry based on INT (In-band Network Telemetry) probes.

In this context, INT telemetry instructions direct the collection of fine-grained metrics on network devices, with low granularity and high frequency, allowing network operators to obtain a clearer picture of the network state from their devices.

However, the existing works in the literature currently focus on collecting performance metrics from network interfaces, assuming that there is only a single queue per output port. However, in more recent programmable devices, there are at least 8 queues associated with each port, and in some more modern equipment, this number can reach up to 128 queues. This number of queues is extremely useful, as different classes of network flows can be used, each queue with a different quality of service.


The main challenge when multiple queues are present is related to the fact that, traditionally, a single probe is capable of collecting telemetry only from the queue it traverses, which is associated with the forwarding output port of the probe. Therefore, if a port has two queues, two probes would need to be sent in order to monitor all those queues. To monitor all the queues of all ports of each switch in the network, a prohibitive increase in probes would be required, raising the monitoring overhead.

In this context, this paper presents \nome{} (Multiqueue Multicast - INT), a solution capable of collecting information from all queues of each physical port of a programmable switch using the P4 language. In summary, the solution developed in this work uses registers to store the metadata of each queue. These registers are updated whenever a data packet passes through a queue in the device. An INT probe, in the \nome{} solution, is capable of collecting the INT metadata from all queues and all ports by reading these registers. The solution proposed in \nome{} contrasts with the traditional INT telemetry-based monitoring mechanism, because instead of obtaining the INT metadata directly from the queue associated with the probe’s output port, it collects the data stored in the device's internal registers. In summary, such registers accumulate monitoring data from all queues associated with all ports of the monitored switch.

Furthermore, the solution presented here makes use of source routing and multipath (multicast) trees to reduce the number of probes required to cover all switches in the network, avoiding overlaps of these probes as they traverse the topology. For this, we adopt the \mpolkaint{} mechanism \cite{mpolkaint}, which uses source-based routing, replacing traditional routing tables with a label used for forwarding packets in the network. The solution was implemented using the P4 language and evaluated in a Mininet environment with BMv2 (Behavioral Model version 2) switches. The evaluation of \nome{} showed that the solution reduced the number of probes sent by 4x and the amount of data transferred for network telemetry collection by approximately 2.8x when compared to traditional solutions found in the literature.




%% file: sections/related.tex
\section{Related Work} \label{sec:relacionados}


This section presents some relevant works from the literature that address in-band network telemetry using INT. More specifically, to highlight the main contributions of this paper, the related work is divided into two categories: (i) use of INT with source routing solutions; (ii) use of INT in devices that have multiple queues or in scenarios that require monitoring at the port-level granularity.

It can be stated that INT-Path~\cite{ref:intpath} was the pioneering work that combined source routing and INT telemetry to enable the specification of a path, represented as a list of nodes included in the packet header. INT-Probe~\cite{ref:intprobe} addresses the same problem but additionally proposes a probe path planning algorithm to achieve full network coverage. SR-INT~\cite{ref:srint} uses a fixed-length header and mitigates the overhead problem of INT probes by leveraging the Segment Routing protocol. Although there is a significant amount of relevant work that incorporates source routing solutions with in-band telemetry using INT, all the cited works so far make use of unicast probes and focus on reducing the monitoring task cost by minimizing the number of emitted INT probes.

With the same goal, MPINT~\cite{ref:mpint} introduces an innovation by applying telemetry using INT to multicast traffic without relying on source routing solutions. In turn, MPolKA-INT~\cite{mpolkaint} is the pioneering work in applying in-band telemetry to multicast traffic using source routing solutions, whose results demonstrated the efficiency of the approach in increasing network coverage with the minimum number of probes.

However, none of these related works address the issue of monitoring multiple queues on the same port. 
In~\cite{ref:virtqueue}, a solution is presented for the use of virtual queues in the P4 pipeline, exploring its application for traffic management in different programmable devices. 
P4-CoDel~\cite{ref:p4-codel}, in turn, highlights challenges and demonstrates, using P4, how to implement Active Queue Management (AQM) algorithms in programmable hardware. The authors also refer to a version of the \texttt{fq\_CoDel} algorithm that addresses the management of multiple queues, but no further details are provided in the article. Finally, Vogt et al.~\cite{ref:ipg} present IPGNET, a monitoring solution based on INT, where the authors highlight the feasibility of correlating telemetry data even across multiple queues. However, IPGNET does not mention the data collection method for those multiple queues.

In summary, as shown in Table \ref{tab:related}, it is easy to identify a gap in the literature regarding the proposition of solutions aimed at: (i) enabling the collection of telemetry data from multiple queues in programmable devices, and (ii) using source routing solutions with \textit{multipath} probes to minimize the overhead in covering the nodes in a P4 programmable network.

\begin{table}[]
\centering \caption{Summary of information on related works.}
\vspace{0.2cm}
\footnotesize	
\begin{tabular}{lcccc}
\hline
\label{tab:related}
\textbf{Ref.} & \textbf{INT} & \textbf{Source Routing} & \textbf{\textit{Multipath}} & \textbf{Multiple Queues}\\
\hline
\cite{ref:intpath} & \textcolor{green}{\checkmark} & \textcolor{green}{\checkmark} & \textcolor{red}{X} & \textcolor{red}{X}\\
\cite{ref:intprobe} & \textcolor{green}{\checkmark} & \textcolor{green}{\checkmark} & \textcolor{red}{X} & \textcolor{red}{X}\\
\cite{ref:srint} & \textcolor{green}{\checkmark} & \textcolor{green}{\checkmark} & \textcolor{red}{X} & \textcolor{red}{X}\\
\cite{ref:mpint} & \textcolor{green}{\checkmark} & \textcolor{red}{X} & \textcolor{green}{\checkmark} & \textcolor{red}{X}\\
\cite{mpolkaint} & \textcolor{green}{\checkmark} & \textcolor{green}{\checkmark} & \textcolor{green}{\checkmark} & \textcolor{red}{X}\\
\cite{ref:virtqueue} & \textcolor{red}{X} & \textcolor{red}{X} & \textcolor{red}{X} & \textcolor{green}{\checkmark}\\
\cite{ref:p4-codel} & \textcolor{red}{X} & \textcolor{red}{X} & \textcolor{red}{X} & \textcolor{green}{\checkmark}\\
\cite{ref:ipg} & \textcolor{green}{\checkmark} & \textcolor{red}{X} & \textcolor{red}{X} & \textcolor{green}{\checkmark}\\
\nome & \textcolor{green}{\checkmark} & \textcolor{green}{\checkmark} & \textcolor{green}{\checkmark} & \textcolor{green}{\checkmark}\\
\hline

\end{tabular}
\end{table}

%% file: sections/conclusao.tex
\section{Conclusions and Future Work} \label{sec:conclusoes}

In this article, we present a solution for telemetry collection in programmable switches, \nome, which enables the collection of data from multiple queues, reducing network overhead and eliminating probe redundancy. \nome{} uses a routing infrastructure at the source and exploits multiple paths to achieve complete coverage of the monitored network. As main results, \nome{} was able to minimize the number of probes required to collect telemetry data from multiple queues using INT, as well as reduce the total Bytes used to cover the collection from all switches in the evaluated topology. Additionally, the solution was incorporated into Mininet and made publicly available.


As future work, the main demands are related to the continuation of the implementation of \nome{}, addressing the limitations described in Section \ref{sec:limit}. Furthermore, it is intended to evaluate the feasibility and performance scalability of the proposal on real programmable switches, such as Tofino 2.

%% file: sections/ack.tex
\section*{Acknowledgments}

The authors thank the financial support from FAPESP, processes $2022/13544-8$ and $2020/05182-3$, and from FAPES, processes $941/2022$, $2023/RWXSZ$ and $2022/ZQX6$.

%% file: main.bbl
\begin{thebibliography}{}

\bibitem[Arslan and McKeown 2019]{McKeown:2019}
Arslan, S. and McKeown, N. (2019).
\newblock Switches know the exact amount of congestion.
\newblock In {\em Proc. of the Workshop on Buffer Sizing}, BS ’19, New York, NY, USA. ACM.

\bibitem[Ben~Basat et~al. 2020]{PINT}
Ben~Basat, R. et~al. (2020).
\newblock Pint: Probabilistic in-band network telemetry.
\newblock In {\em Proceedings of the ACM SIGCOMM}, SIGCOMM '20, page 662–680, New York, NY, USA. ACM.

\bibitem[de~O.~Pereira et~al. 2023]{mpolkaint}
de~O.~Pereira, I. et~al. (2023).
\newblock {MPolKA-INT: Stateless Multipath Source Routing for In-Band Network Telemetry}.
\newblock In Barolli, L., editor, {\em Advanced Information Networking and Applications}, pages 513--524, Cham. Springer International Publishing.

\bibitem[Guimarães et~al. 2022]{mpolka}
Guimarães, R.~S. et~al. (2022).
\newblock {M-PolKA: Multipath Polynomial Key-Based Source Routing for Reliable Communications}.
\newblock {\em IEEE Transactions on Network and Service Management}, 19(3):2639--2651.

\bibitem[Harkous et~al. 2021]{ref:virtqueue}
Harkous, H. et~al. (2021).
\newblock Virtual queues for p4: A poor man’s programmable traffic manager.
\newblock {\em IEEE Transactions on Network and Service Management}, 18(3):2860--2872.

\bibitem[{Kim} et~al. 2018]{Kim:2018}
{Kim}, Y. et~al. (2018).
\newblock Buffer management of virtualized network slices for quality-of-service satisfaction.
\newblock In {\em IEEE Conf on Network Function Virtualization and Software Defined Networks (NFV-SDN)}, number 18725013 in 1, pages 1--4, Verona, Italy. IEEE.

\bibitem[Kundel et~al. 2021]{ref:p4-codel}
Kundel, R. et~al. (2021).
\newblock P4-codel: Experiences on programmable data plane hardware.
\newblock In {\em ICC 2021 - IEEE International Conference on Communications}, pages 1--6.

\bibitem[Pan et~al. 2019]{ref:intpath}
Pan, T. et~al. (2019).
\newblock Int-path: Towards optimal path planning for in-band network-wide telemetry.
\newblock In {\em IEEE INFOCOM 2019}, pages 487--495. IEEE.

\bibitem[Pan et~al. 2021]{ref:intprobe}
Pan, T. et~al. (2021).
\newblock Int-probe: Lightweight in-band network-wide telemetry with stationary probes.
\newblock In {\em IEEE 41st ICDCS}, pages 898--909.

\bibitem[Papadopoulos et~al. 2023]{DINT-PINT}
Papadopoulos, K., Papadimitriou, P., and Papagianni, C. (2023).
\newblock Deterministic and probabilistic p4-enabled lightweight in-band network telemetry.
\newblock {\em IEEE Transactions on Network and Service Management}, 20(4):4909--4922.

\bibitem[Sharma et~al. 2015]{QueueDisciplines}
Sharma, R., Sehra, S., and Sehra, S.~K. (2015).
\newblock Review of different queuing disciplines in voip, video conferencing and file transfer.
\newblock {\em IJARCCE}, pages 264--267.

\bibitem[Tang et~al. 2020]{SEL-INT}
Tang, S., Li, D., Niu, B., Peng, J., and Zhu, Z. (2020).
\newblock Sel-int: A runtime-programmable selective in-band network telemetry system.
\newblock {\em IEEE Transactions on Network and Service Management}, 17(2):708--721.

\bibitem[Vogt et~al. 2022]{ref:ipg}
Vogt, F.~G. et~al. (2022).
\newblock Innovative network monitoring techniques through in-band inter packet gap telemetry (ipgnet).
\newblock In {\em Proc of the 5th Int Workshop on P4 in Europe}, EuroP4 '22, page 53–56, New York, NY, USA. ACM.

\bibitem[Zheng et~al. 2021a]{ref:srint}
Zheng, Q., Tang, S., Chen, B., and Zhu, Z. (2021a).
\newblock Highly-efficient and adaptive network monitoring: When int meets segment routing.
\newblock {\em IEEE Transactions on Network and Service Management}, 18(3):2587--2597.

\bibitem[Zheng et~al. 2021b]{ref:mpint}
Zheng, Y., Pan, T., Zhang, Y., Song, E., Huang, T., and Liu, Y. (2021b).
\newblock Multipath in-band network telemetry.
\newblock In {\em IEEE INFOCOM 2021 - IEEE Conference on Computer Communications Workshops (INFOCOM WKSHPS)}, pages 1--2.

\end{thebibliography}
